\documentclass{PoS}
\pdfoutput=1
\usepackage[utf8]{inputenc}
\usepackage{amsmath,amssymb}
\usepackage{graphicx}
\usepackage{graphicx}
\usepackage{picture}

\usepackage[sort&compress,numbers,merge]{natbib}
\usepackage{natbib,natbibspacing}

\title{Discretization Errors for
the Gluon and Ghost Propagators in Landau Gauge using NSPT}

\ShortTitle{Discretization Errors for the Gluon and Ghost Propagators using
NSPT}

\author{\speaker{Jakob Simeth}\\%
        Institut f\"ur Theoretische Physik, Universit\"at Regensburg, 93040
Regensburg, Germany\\
        E-mail: \email{jakob.simeth@physik.uni-regensburg.de}}

\author{Andr\'e Sternbeck\\%
        Institut f\"ur Theoretische Physik, Universit\"at Regensburg, 93040
Regensburg, Germany\\
        E-mail: \email{andre.sternbeck@physik.uni-regensburg.de}}

\author{Ernst-Michael Ilgenfritz\\
    Joint Institute for Nuclear Research, VBLHEP, 141980 Dubna, Russia}

\author{Holger Perlt\\
  Institut f\"ur Theoretische Physik, Universit\"at Leipzig, 04109 Leipzig,
  Germany}
  
\author{Arwed Schiller\\
  Institut f\"ur Theoretische Physik, Universit\"at Leipzig, 04109 Leipzig,
Germany}

\abstract{
The subtraction of hypercubic lattice corrections, calculated at 1-loop 
order in lattice perturbation theory (LPT), is common practice, e.g., for 
determinations of renormalization constants in lattice hadron physics. 
Providing such corrections beyond 1-loop order is however very demanding in LPT,
and numerical stochastic perturbation theory (NSPT) might be the better 
candidate for this. Here we report on a first feasibility check of this 
method and provide (in a parametrization valid
for arbitrary lattice couplings) the lattice corrections up to 3-loop order for
the SU(3) gluon and ghost propagators in Landau gauge. These propagators are
ideal candidates for such a check, as they are available from lattice
simulations to high precision and can be combined to a renormalization group
invariant product (Minimal MOM coupling) for which a 1-loop LPT correction was
found to be insufficient to remove the bulk of the hypercubic lattice artifacts
from the data. As a bonus, we also compare our results with the ever popular
H(4) method.
}

\FullConference{31st International Symposium on Lattice Field Theory LATTICE 2013\\
		 July 29 – August 3, 2013\\
		 Mainz, Germany}

\begin{document}

\section*{Introduction}
With the advancement of computer technologies, lattice QCD calculations have
become an essential tool to tackle problems of strong interaction physics. For
many applications nowadays the statistical uncertainties have been reduced
to such a degree that a full understanding of all systematic
errors is more important to reach high precision than a further increase of
statistics. 

One of these systematic errors comes from the lattice itself: The symmetry
group on the lattice is the hypercubic group $H(4)$, a subgroup of $O(4)$. 
This introduces lattice artifacts for any lattice momentum but zero
which only disappear in the continuum limit. Of course, the severity of
these hypercubic lattice artifacts depends on the lattice spacing $a$, but since
lattice calculations are currently done for values of $a$ between
$0.04$ and $0.09\,\text{fm}$ it may be important to know these artifacts for
one's favorite observable on a quantitative rather than qualitative level.
In general, discretization effects are most visible at large lattice momentum
\(a^2p^2\) with
$p_\mu = 2\pi k_\mu/ aN_\mu$ \big(with $k_\mu\in(-N_\mu/2,N_\mu/2]$\big).
However, this dependence is non-monotonous as
lattice observables transform as
functions of the four \(H(4)\)
invariants \(\{p^2, p^{[4]}, p^{[6]}, p^{[8]}\}\)
with \(p^{[n]} \equiv \sum_\mu p_\mu^n\) rather than just of $p^2$ as in the
continuum.

There are mainly two approaches that have been followed in the past to
treat these artifacts. One is the so-called $H(4)$-method (see, e.g.,
\cite{Becirevic:1999uc,*Becirevic:1999hj,*Boucaud2003,*Soto2007}), which relies
on fitting coefficients of an expansion in the hypercubic invariants. 
Another is calculating hypercubic corrections in lattice perturbation theory (LPT).
This has been used and developed for precision determinations of renormalization 
constants for hadronic operators
\cite{Gockeler:2010yr,*Constantinou:2013ada,*Constantinou:2013ova}
where an efficient treatment of such artifacts is crucial.
For these calculations one often uses plane-wave sources for the inversion of the
fermion matrix to keep the statistical noise at a minimum and also restricts
to momenta close to the lattice diagonal to minimize hypercubic lattice
artifacts.
To further reduce discretization errors, one subtracts the
difference $\Delta O(p)$ of an operator $O$ in LPT at finite lattice spacing~$a$
and its value in the corresponding limit $a\to0$. 
Unfortunately, 
the calculation of \(\Delta O(p)\) even at 1-loop order can be rather involved in
LPT for certain operators and actions.
Furthermore, in some cases a 1-loop correction does not
suffice.

Numerical stochastic lattice perturbation theory (NSPT) provides a way
out: It allows for a non-diagrammatic but stochastic treatment of LPT
to high loop orders and so to get an estimate for $\Delta O(p)$ beyond
1-loop.\footnote{For a NSPT calculation of renormalization constants 
see for example \cite{Brambilla:2013sua}.}

We will test the feasibility and precision of such an approach here for the
gluon~($Z$) and ghost~($J$) dressing functions and in particular for the Minimal
MOM coupling in Landau gauge~\cite{vonSmekal:1997is,*vonSmekal:2009ae}
\begin{equation}
    \alpha_s^{\mathsf{MM}}(p) = \frac{g_0^2(a)}{4\,\pi} Z(a, p) J^2(a,
p)\text{,}\qquad a \rightarrow 0\text{,}
    \label{eq:minimom}
\end{equation}
which serves as an ideal benchmark for this: Its renorma\-lization-group
invariance is slightly broken on the lattice at larger momenta by the
hypercubic lattice artifacts of the gluon and ghost dressing
functions and a subtraction of the exact 1-loop correction was
shown to be insufficient to repair this in the data for
$\alpha_s^{\mathsf{MM}}$~\cite{Sternbeck2012}. It also allows us to
focus first on quenched QCD, keeping the additions of fermions to
the NSPT calculations for later.
%
%
\section*{$Z$ and $J$ from Numerical Stochastic Perturbation Theory}
\begin{figure*}
\centering
\mbox{\includegraphics[height=6cm]{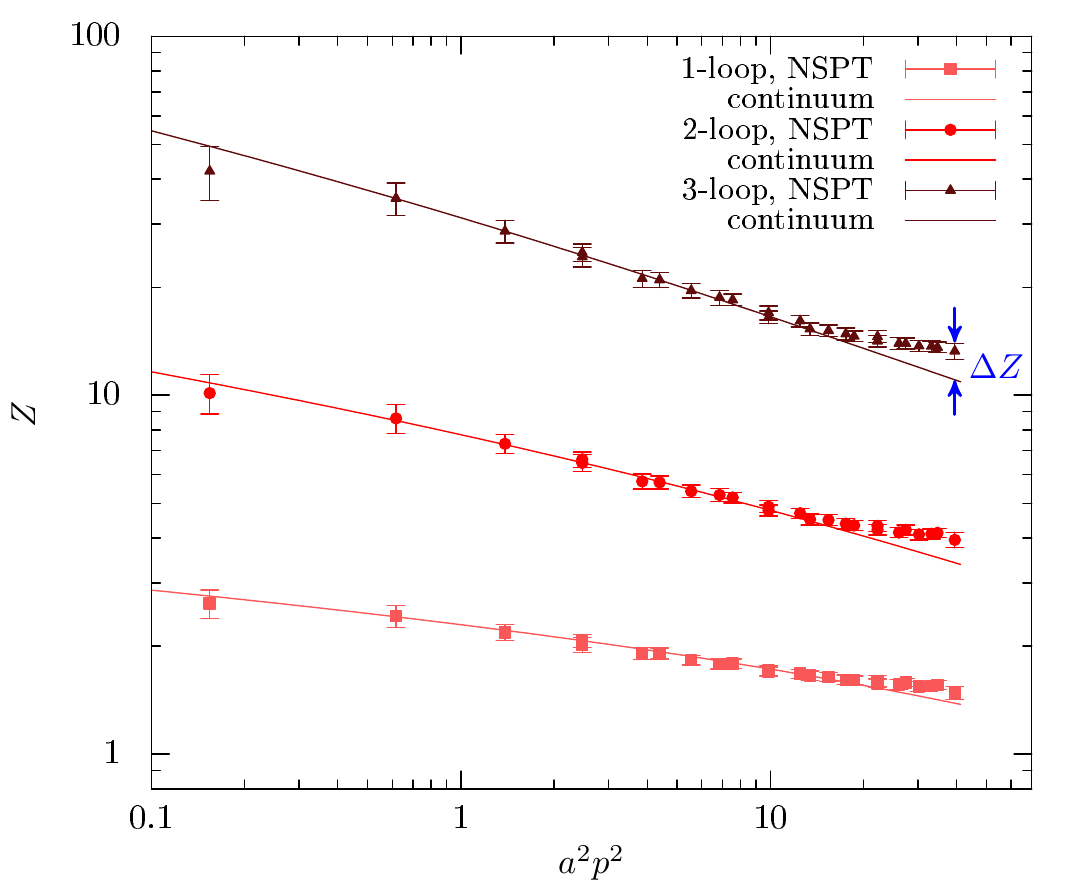}
      \includegraphics[height=6cm]{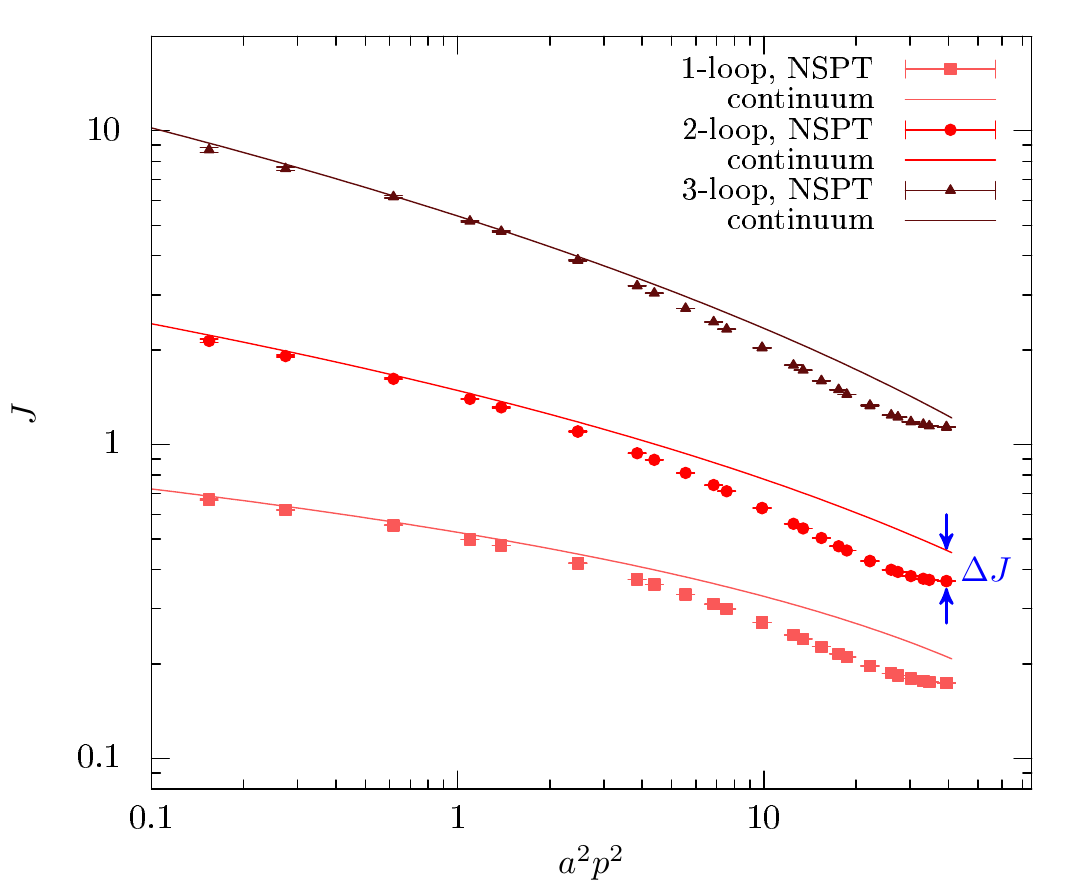}}
\caption{NSPT data for the gluon (left) and ghost (right) dressing 
functions for
diagonal lattice momenta up to 3-loop order. Curves represent the $a\to0$
result (see text). Arrows mark the hypercubic lattice artifacts.}
\label{fig:hypcor}
\end{figure*}

To get the desired hypercubic corrections, we have to calculate the Landau-gauge
gluon and ghost propagators in
NSPT. Details for this can be found in \cite{DiRenzo:2009ni,*DiRenzo2010cs}.
Here it suffices to mention that their tensor structure (for the Wilson
gauge action) is of the familiar form
\begin{equation}
    D_{\mu\nu}^{ab} =
\left(\delta_{\mu\nu}-\frac{\hat{p}_\mu\,\hat{p}_\nu}{\hat{p}^2}\right)\delta^{
ab} \frac{Z}{\hat{p}^2}\qquad\text{and}\qquad
    G_{\mu\nu}^{ab} = \delta_{\mu\nu}\delta^{ab}\frac{J}{\hat{p}^2}
    \label{eq:dressingfn}
\end{equation}
where \(\hat{p}_\mu = \frac{2}{a}\sin(a\,p_\mu/2)\) denotes the eigenvalues
of the free lattice Laplacian.
The results are obtained in orders of \(g_0^2\), because the \(SU(3)\) link
fields \(U_{x,\mu}\) are kept as an expansion in powers of \(\beta=2N_c/g_0^2\) which
also requires algebraic operations to be performed with respect to this 
perturbative structure for operators $A$ and $B$:
\begin{equation}
    U_{x,\mu} = 1 + \sum_{k \ge 1} \beta^{-k/2} U_{x,\mu}^{(k)}\text{,}\qquad
    (A+B)^{(k)} = A^{(k)}+B^{(k)}\text{,}\qquad (A\,B)^{(k)} = \sum_{j=0}^k
A^{(k)}\,B^{(k-j)}\text{.}
\end{equation}
Furthermore, the framework of NSPT relies on stochastic quantization. This
is implemented by adding an artificial time $\tau$ to the gauge fields whose
evolution in $\tau$ is governed by a Langevin equation, which is
solved numerically. A suitable Euler scheme reads
\begin{equation}
    U_{x,\mu}(\tau+\epsilon) = \exp\{-i\,\epsilon\,\nabla_{U_{x,\mu}}S[U] +
\sqrt{\epsilon}\,\eta_x(\tau)\}\,U_{x,\mu}(\tau)\text{,}
\end{equation}
where \(\eta = \eta^a t^a\) is a random gaussian distributed \(su(3)\) field
(\(t^a\) being the generators of \(SU(3)\)) and \(S[U]\) (in our case) the Wilson plaquette
action. \(\nabla_{U_{x,\mu}}\) is the discretized left Lie derivative with
respect to \(U_{x,\mu}\). Since the Euler time step $\epsilon$ has to be finite,
simulations have to be performed for several \(\epsilon\) such that at the end
the limit \(\epsilon\rightarrow 0\) can be taken.

We choose $\epsilon=0.03$, 0.02 and 0.01 for all our simulations and
collect data for $J^{(l)}$ and $Z^{(l)}$ for $l=1,\ldots,4$ loops after fixing 
$U_{x,\mu}$ to Landau gauge. We will
focus here on diagonal lattice momenta and our lattice sizes are $8^4$, $16^4$,
$24^4$ and $32^4$. Data for a $48^4$ lattice is in progress.

In Figure~\ref{fig:hypcor} we show our NSPT data for the gluon and ghost
dressing functions up to 3-loop order. As we are interested in the hypercubic
lattice corrections for diagonal lattice momenta we only show points for these
momenta and compare it to the momentum dependence 
\begin{equation}
    J^{(i)}_{cont}(a^2\,p^2) = j^{(i)}_{0} + \sum_{k=1}^{i}
j^{(i)}_{k}\left(\log(a^2\,p^2)\right)^k\text{.}
\end{equation}
expected in the limit $a\to0$ (for simplicity we write \(J\) meaning both
\(J\) and \(Z\)). For the 1-loop curve we use the well-known coefficients,
$j^{(1)}_0$ and $j^{(1)}_{1}$, from \cite{Kawai:1980ja}, while for the 2-loop
and 3-loop curves we use the coefficients $j^{(i=2,3)}_{k>0}$ multiplying the
divergent logarithms from 3-loop continuum perturbation theory
\cite{Gracey2003}, and the finite \(j^{(i)}_{0}\) from our fits (see below).
Within errors our values for the latter agree with those found
in \cite{DiRenzo:2009ni,*DiRenzo2010cs}. For a better illustration, we
have marked the momentum-dependent hypercubic lattice corrections $\Delta Z$ and
$\Delta J$ in Figure~\ref{fig:hypcor} which we aim to quantify:
\begin{equation}
   \Delta J^{(i)} =  J^{(i)}_{lat}(a^2p^{2}, a^4p^{[4]}, a^6p^{[6]}, a^8p^{[8]})
    - J^{(i)}_{cont}(a^2\,p^2) \text{.}
    \label{eq:hypcor}
\end{equation}\vspace{-4ex}

\begin{figure*}
  \mbox{\includegraphics[height=6cm]{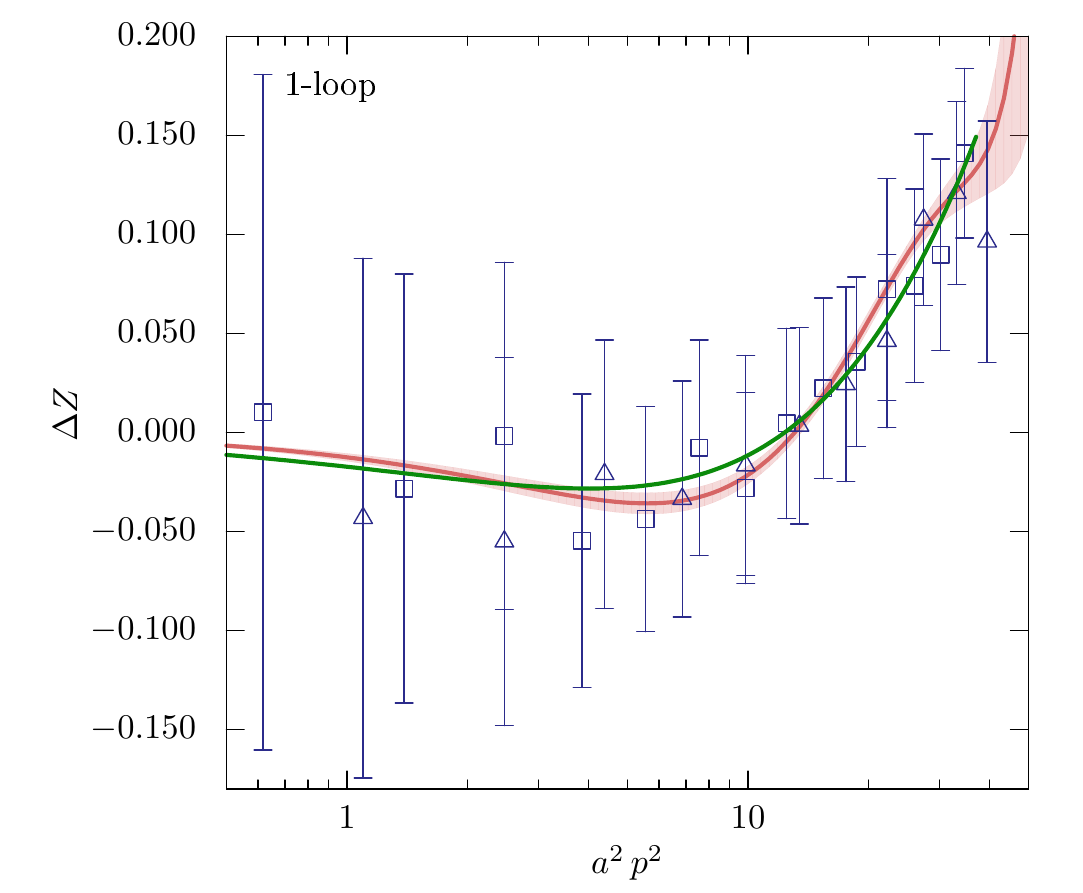}
        \includegraphics[height=6cm]{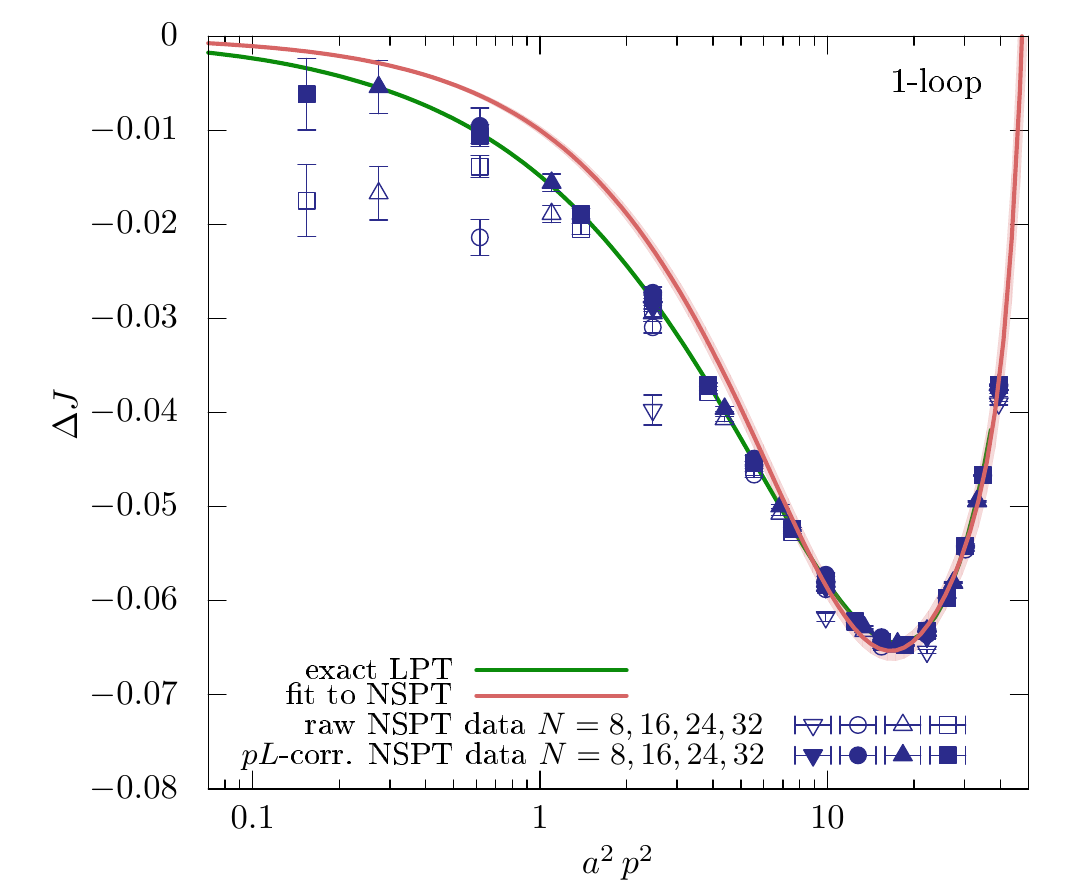}}\vspace*{-1ex}
        \caption{Hypercubic lattice artifacts at 1-loop order for the gluon 
            (left) and ghost (right) dressing functions. $pL$-corrected
  (uncorrected) data is shown by full (open) symbols. Green curves are the
  exact 1-loop LPT results from~\cite{Sternbeck2012}. Red bands are from a fit
to the data.}
  \label{fig:lptnspt}
  \medskip
 \mbox{\includegraphics[height=6cm]{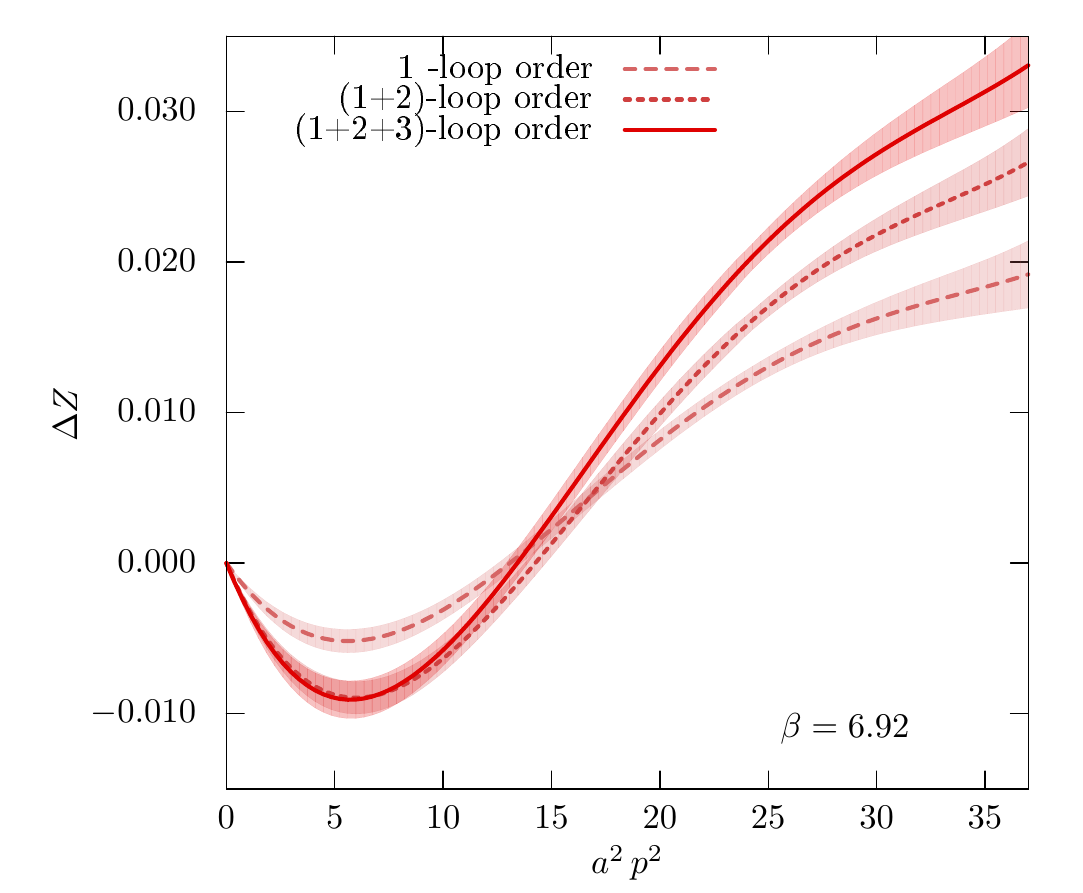}
        \includegraphics[height=6cm]{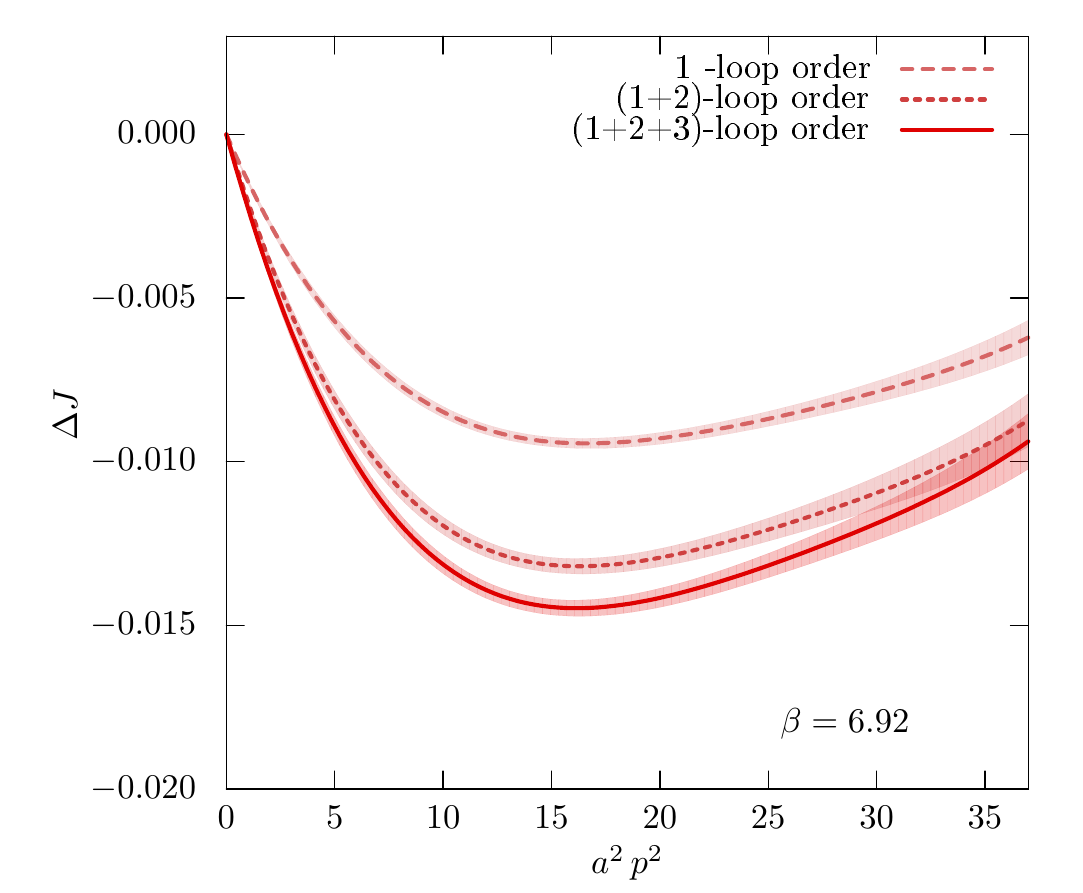}}\vspace*{-1ex}
 \caption{Hypercubic lattice artifacts up to 3-loop order 
for the gluon (left) and ghost (right) dressing functions at \(\beta=6.92\). 
Bands result from a fit to our NSPT data.}
 \label{fig:glughocor}
\end{figure*}

\section*{Removal of finite size effects}
Although a first look at Figure~\ref{fig:hypcor} suggests we will get
$\Delta Z$ and $\Delta J$ straightaway, a closer inspection of
the raw data reveals we have to carefully treat finite size effects 
first. These are visible in particular for the ghost propagator at
small momenta (see, e.g., the open symbols in Figure~\ref{fig:lptnspt}).

To treat these effects we follow a procedure proposed in
\cite{DiRenzo:2009ni,*DiRenzo2010cs}, albeit in a slightly different manner.
It starts with the observation that an observable \(O\) on the lattice only
depends on dimensionless quantities, namely on $ap$ (in the way as mentioned
above) and on $pL$ which comes in due to the finite lattice extend $N=L/a$:
\vspace{-1ex}
\begin{equation}
    O(p\,a, p\,L) = O(p\,a, \infty) + \delta O(p\,a,p\,L)\text{.}
\end{equation}
To get $O$ in the infinite volume limit one can either extrapolate data
for several volumes or calculate and subtract $\delta O(p\,a,p\,L)$ from the
data. We choose the latter, which is more robust in our case, and determine
\(\delta O(p\,a,p\,L)\) in the following way, similarly to \cite{DiRenzo:2009ni,*DiRenzo2010cs}:
We assume we can neglect the influence of additional (hypercubic)
\(p\,a\)-corrections to $\delta O$, 
i.e., \(\delta O(p\,a, p\,L) \approx \delta O(0, p\,L)\). 
This quantity then only depends on the integer tuple \(k\) that defines the
momentum $ap$, because
\(    p_\mu\,L = \frac{2\,\pi\,k_\mu}{N\,a} N\,a = 2\,\pi\,k_\mu\text{.}\)
Furthermore, an observable \(O(pa,\infty)\) is expected to be equal for equal
physical momenta, 
i.e., for equal \(k/N\). That is, the difference must be due to
the finite-size effect and we can write:
\begin{enumerate}
    \item \(\delta O(k, N_1) = \delta O(k, N_2)\)
    \item \(\delta O(k_1, N_1) = O(k_1, N_1) - \left[O(k_2, N_2) - \delta O(k_2,
        N_2)\right]\) \qquad \text{if}\quad \(k_1/N_1 = k_2/N_2\)\,.
\end{enumerate}
This allows us to go through our data along sequences of links (paths), each
connecting two data points of either constant \(k/N\) or \(k\), and to get
the \(p\,L\)-error for each path recursively. As a starting point for
this recursion we choose \(\delta O(k_{max}, N_{max})=0\) which should be valid
for large \(N_{max}\).

By that procedure we can reach nearly all pairs of \((k, N)\). Most of them can 
be accessed by many different paths over which we average. The \(pL\)-error of
points that are disconnected from the anchor \((k_{max}, N_{max})\) is
interpolated linearly between neighbouring values of \(k/N\).

This method of finite-size corrections turns out to be very successful. To
demonstrate that we show in Figure \ref{fig:lptnspt} (right) our
1-loop data for the uncorrected (open symbols) and the $pL$-corrected (full
symbols) difference $\Delta J$ (Eq.~\eqref{eq:hypcor}). There we see, finite
size errors have quite an effect at low momenta and it is necessary to remove
them, but after the $pL$-correction the $\Delta J$ data from different
lattices falls on top of the curve one knows exactly from 1-loop LPT (from
\cite{Sternbeck2012}). 
\begin{figure*}
\vspace*{-0.5cm}
\centering
\includegraphics[height=5.7cm]{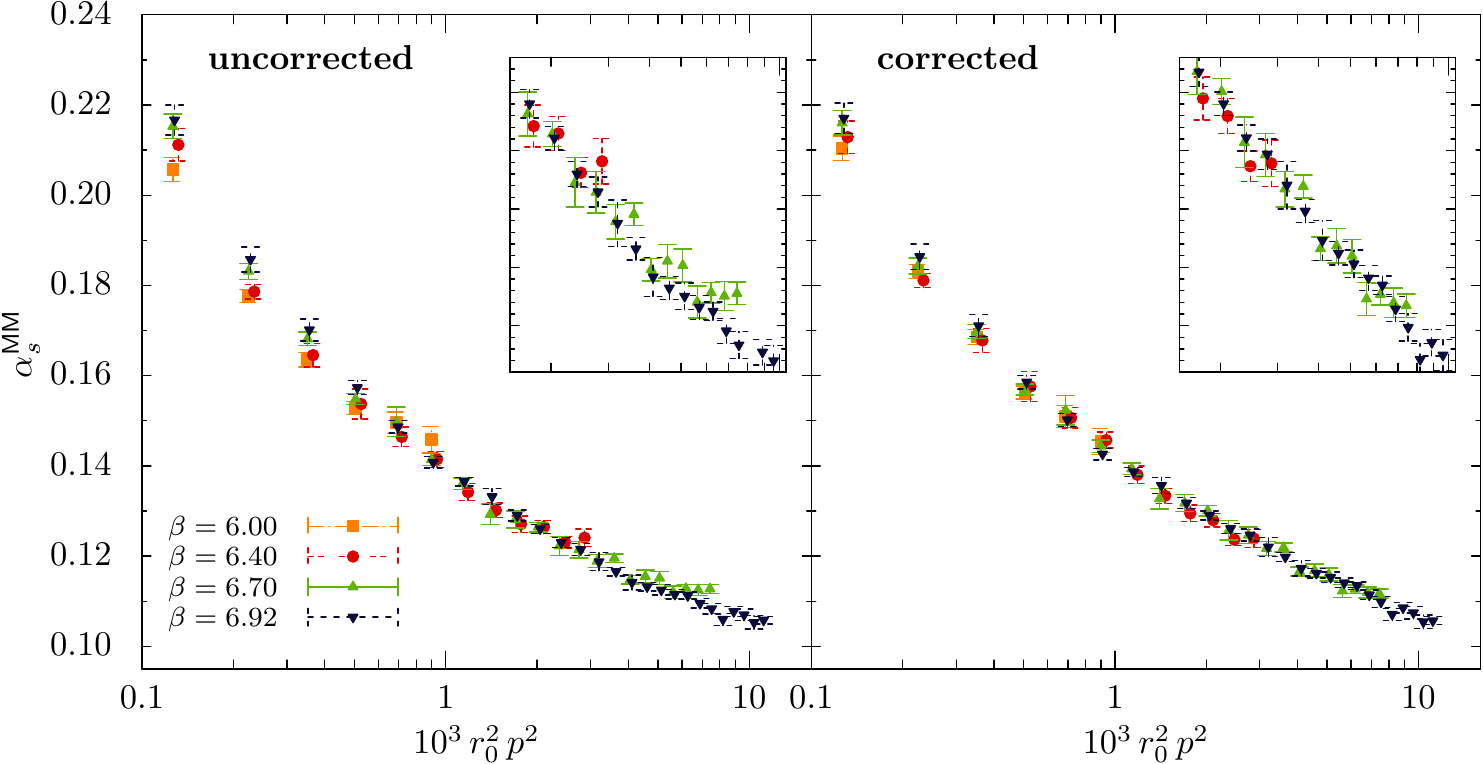}
\caption{Coupling constant in the Minimal MOM scheme for quenched QCD in
Landau gauge. Data is for a fixed physical volume (from \cite{Sternbeck2012}),
before (left) and after (right) correcting for discretization
effects.}
\label{fig:coupling}
\end{figure*}
The gluon propagator data is currently too noisy for a removal of $pL$-effects
and it thus remains uncorrected in what follows. Though, within errors,
points lie on the 1-loop LPT curve already.

\section*{Results}
Now, that we have checked that our 1-loop NSPT results conform with the exact
from LPT, we can quantify the hypercubic lattice artifacts
(Eq.~\eqref{eq:hypcor}) up to 3-loop order. To ease their later use we will
provide them in a parametrization independent of the lattice coupling
\(\beta\). For this we fit each order separately to a polynomial of the \(H(4)\)
invariants:
\begin{equation}
    \Delta J^{(i)}(p^{2}, p^{[4]}, p^{[6]}, p^{[8]})=c_2\,p^2 + c_4\,p^{[4]} +
c_6\,p^{[6]} + c_8\,p^{[8]}\text{.}
\end{equation}
The (summed-up) hypercubic lattice artifacts are then obtained for each
value of $\beta$ from:
\begin{equation}
\Delta J(\beta,p^{2}, p^{[4]}, p^{[6]}, p^{[8]}) = \sum_{i=1}^{3} \beta^{-i}
\Delta J^{(i)}(p^{2}, p^{[4]}, p^{[6]}, p^{[8]})\text{.}
\end{equation}

A comparison with the exact 1-loop LPT result from \cite{Sternbeck2012} shows
that these fits describe the overall momentum dependence quite well, even
though it cannot describe it perfectly (see Figure~\ref{fig:lptnspt}). There
the fit is dominated by the many points at higher momentum while being
constrained to zero at \(a^2p^2=0\). So there are small deviations for small
$a^2p^2$. We therefore regard this fit model not as best but as best usable
to parametrize the leading hypercubic artifacts.

The final results for $\Delta Z$ and $\Delta J$ can be seen in
Figure~\ref{fig:glughocor}. There we choose $\beta\equiv 6.92$ and one sees
that even for this fine lattice (\(a\approx 0.026\,\text{fm}\)) the 3-loop 
order correction still contributes rather significantly to the gluon 
dressing function. The ghost dressing function needs improvement to at 
least 2-loop order. 
This explains why a 1-loop correction of the data for
$\alpha_s^{\mathsf{MM}}$ is insufficient as seen in \cite{Sternbeck2012}. Our
new results allows us now to correct the hypercubic lattice artifacts up to
3-loop order, and, as evidenced in Figure~\ref{fig:coupling}, with these we
are successful: The corrected $\alpha_s^{\mathsf{MM}}$ data (from
\cite{Sternbeck2012}) for the different lattice spacings falls on top of each
other for \emph{all} momenta. Renormalization-group invariance is thus
restored.
%
%
\section*{Comparison with the H(4)-Method}
\begin{floatingfigure}[r]
  \centering
  \vspace*{-0.2cm}
  \parbox{6.7cm}{%
    \includegraphics[width=7.cm]{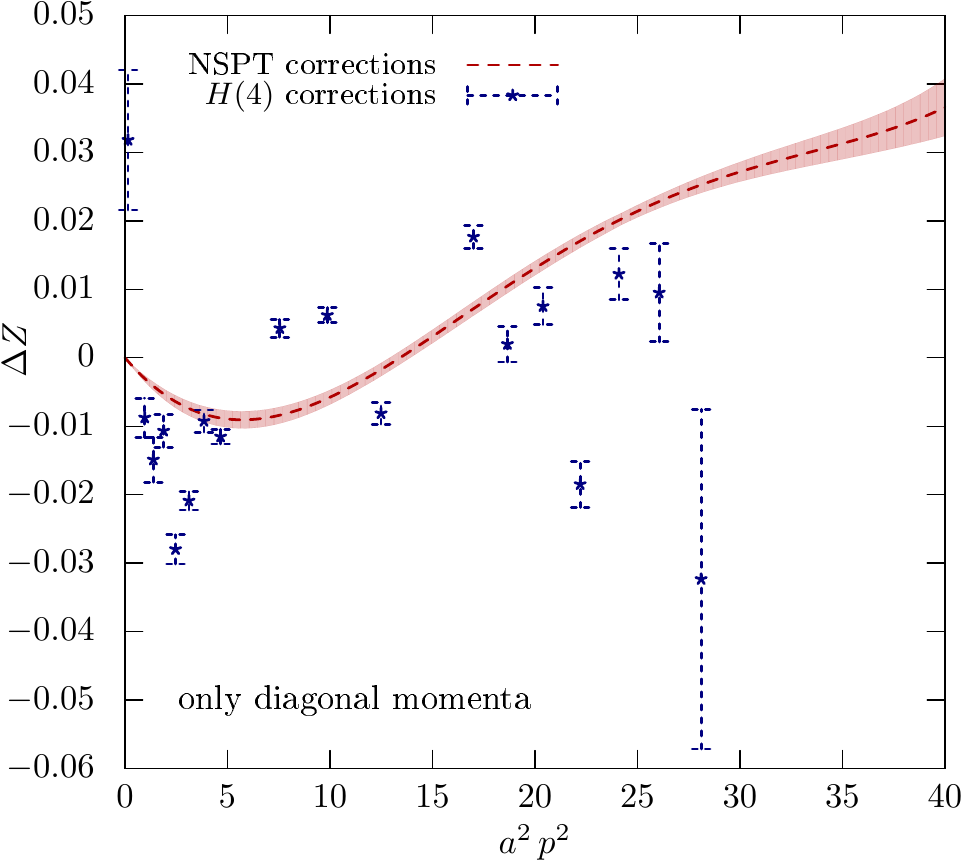}}
    \caption{Hypercubic corrections for the gluon dressing function
    for diagonal momenta ($64^4$ lattice at \(\beta=6.92\)).
    Blue crosses are from the \(H(4)\)-method, using all
    available orbits (not shown). The curve is the 3-loop
    correction from NSPT.
    \label{fig:h4method}\vspace*{-1cm}}
\end{floatingfigure}
It is interesting to compare our approach to the above-mentioned 
\(H(4)\)-method. With this, one tries to fit the coefficients \(c_i\) of
the hypercubic expansion of $O(ap)$, e.g., truncated at
\(\mathcal{O}(a^4)\),
\begin{equation}
    O(a\,\hat{p}) = O(a^2\,p^2) + c_2\,a^2\frac{p^{[4]}}{p^2} +
c_4\,a^4\,p^{[4]} + \cdots\text{.}
    \label{eq:h4fit}
\end{equation}
regarding \(O(a^2p^2)\) as the discretization-artifacts-free
operator~\cite{Becirevic:1999uc,*Becirevic:1999hj,*Boucaud2003,*Soto2007}.
A simultaneous fit of all the $c_i$'s in \eqref{eq:h4fit} requires
many degenerate \(H(4)\) or\-bits for nearby $a^2p^2$, that is this method 
works best for intermediate momenta on large lattices. For a fair comparison,
we therefore choose a $64^4$ lattice, for which we have data for $Z$
for many different orbits. Looking again at diagonal momenta, we find the
$H(4)$ correction scatter, where available, around our 3-loop NSPT
correction (see Figure~\ref{fig:h4method}).

%
\section*{Conclusion}
We have developed a new way to determine and remove discretization errors
which are present in any lattice observable due to the hypercubic lattice
symmetry. Our method is based on a perturbative calculation in the framework of
NSPT, and a first application to quenched data for the Minimal
MOM coupling in Landau gauge has been very successful (see
Figure~\ref{fig:coupling}). 

Our method can easily be extended, for instance, to aid calculations of
renormalization constants for hadronic operators. It is planned to test this
next. More details on our approach will be given in a forthcoming article.\par
%
\acknowledgments
We thank Francesco Di Renzo for insights on the determination of finite
size effects. This work is supported by the European Union under the Grant
Agreement IRG 256594. Grants of time on the Linux-Cluster of the Leibniz
Rechenzentrum in Munich (Germany) are acknowledged.

{\small 
%

}

\end{document}